# Amplification of acoustic phonons in superlattice

K.A. Dompreh, S.Y. Mensah, N.G. Mensah, S.S. Abukari
Dept. of Physics, Uni. of Cape coast, Ghana
F.K.A. Allotey,
Institute of Mathematical Sciences
G.K. Nkrumah-Buandoh,
Dept. of Physics, Uni. of Ghana, Legon

**Abstract**
The amplification of acoustic phonons in a superlattice in the presence of an electric field $E = E_0 + E_1 \cos(\omega t)$ has been investigated theoretically and numerically by computational methods. The calculation is done in the hypersound regime $(ql \gg 1)$ where the attenuation coefficient depends on the phonon wave vector $q = [\frac{-\pi}{d}, \frac{2\pi}{d}]$ and the frequency $\omega_q = 10^{12} s^{-1}$. An inversion is attained where amplification far exceeds absorption and the ratio $\frac{|\Gamma/\Gamma_0|_{min}}{|\Gamma/\Gamma_0|_{max}} \approx 3$. A high frequency build up of acoustic energy from noise (phonon spectrum) is obtained by using specialized spectral techniques. This indicates an amplification of the phonons generated in the terahertz range leading to the possibility of obtaining a hypersound MASER.

## Introduction

Semiconductor superlattice has attracted much attention in the last decade [1-10] as the main form of archiving the construction of compact and reliable sources, amplifies and detectors of Terahertz spectral range of electromagnetic radiation [11]. The possibility of either by varying the SL parameters or by external field makes it suitable for opto - and acousto electronic devices. The attainment of the THz oscillator/amplifier was hindered by the destructive high – field domains in the nanostructure but recent technique as such the fabrication of synthetic materials by molecular beam epitaxy (MBE) and metal-organic chemical vapor deposition (MOCVD) technology has made the superlattice much attainable [12 – 15].The technique of considering the effect of high – frequency electric field on the hypersound amplification by Mensah et el [16] is also studied. The field modulates the amplification coefficient which favors a regenerative THz amplifier/oscillator where it is possible to attain a hypersound MASER. In this paper, a modeling technique is employed to develop an algorithm using spectral analysis and specialized toolboxes [17-22] for the absorption coefficient to generate the acoustic phonons. From it, a high frequency acoustic noise is produced at a specified sampling frequency. Photostimulated attenuation is also analyzed [23, 24]. In this paper, the first part deals with the theory of the attenuation coefficient where the equation of the sound absorption is deduced. In the second session, certain spectral methods are applied to the equation to deduce the spectrum, sound for the analysis of the hypersound MASER.

## Calculation

In solving this problem, the quasi-classical case is considered as in [25]. The sound waves and the applied electric field E are propagated along the axis of the z - axis of the Superlattice. The sound absorption coefficient is given by the expression [26]

$$\Gamma = \frac{|\Lambda|^2 q^2}{4\pi^2 \rho s \omega_q} \int [f(\varepsilon_p) - f(\varepsilon_{p+q})] \delta(\varepsilon_{p+q} - \varepsilon_p - \omega_q) d^3 p \qquad (1)$$

where $\Lambda$ is the constant of deformation potential, $\rho$ is the density of the sample, s is the velocity of sound, $f(\varepsilon_p)$ is the distribution function and $p$ is the momentum of the electron with the energy of the Superlattice in the lowest miniband given as

$$\varepsilon(p) = \frac{p_\perp^2}{2m} + \Delta[1 - \cos(p_z d)] \qquad (2)$$

Here $p_\perp$ and $p_z$ are the quasi-momentum components across and along the superlattice axis resp. taking $k=1$ and $\hbar = 1$. The equilibrium distribution function given by

$$f_0(p) = \frac{\pi d n}{m T I_0(\frac{\Delta}{T})} \exp\left(\frac{-\varepsilon(p)}{T}\right) \qquad (3)$$

The kinetic equation used to determined the distribution function is given by the Boltzmann equation in the $\tau$ - approximation as

$$\frac{\partial f(p,t)}{\partial t} + e[E_0 + E_1 \cos(\omega \tau)]\frac{\partial (p,t)}{\partial t} = -\frac{1}{\tau}[f(p,t) - f_0(p)] \qquad (4)$$

solution of which is given as

$$f(p,t) = \int_0^\infty \exp(\frac{-t'}{\tau}) \frac{dt'}{\tau} f_0[p - (eE_0 t' + \frac{eE_1}{\omega}\{\sin(\omega) - \sin[\omega(t-t')]\})] \qquad (5)$$

For a non-degenerate electron gas, the absorption coefficient from Eq (1) is solved to obtain

$$\Gamma = \Gamma_0 \int_0^\infty \exp(\frac{-t'}{\tau}) \frac{dt'}{\tau} \times \left\{ \sinh\left[\frac{\omega_q}{2T} \cos(eE_0 t' + \frac{eE_1}{\omega}\{\sin(\omega t) - \sin[\omega(t-t')]\})\right] \right.$$

$$\times \cosh\left[\frac{\Delta}{T} \cos(\frac{1}{2}qd)\cos(eE_0 t' + \frac{eE_1}{\omega}\{\sin(\omega t) - \sin[\omega(t-t')]\})\sqrt{1-b^2}\right]$$

$$- \left\{\frac{\Delta}{T}\sqrt{1-b^2}\sin(eE_0 t' + \frac{eE_1}{\omega}\{\sin(\omega t) - \sin[\omega(t-t')]\})\right.$$

$$\left.\times \sin(\frac{1}{2}qd)\sinh[\frac{\Delta}{T}\cos(\frac{1}{2}qd)(eE_0 t' + \frac{eE_1}{\omega}\{\sin(\omega t) - \sin[\omega(t-t')]\})\sqrt{1-b^2}\right\} \qquad (6)$$

where $\Gamma_0 = \frac{|\Lambda|^2 q^2 n \Theta(1-b^2)}{2\rho s \omega_q \Delta \sin(qd/2)\sqrt{1-b^2}}$ and the $\Theta$ is the Heaviside step function and $b = \frac{\omega_q}{2\Delta \sin(qd/2)}$. Finally, by making $T \gg \Delta, \omega_q$ and integrating by averaging over the period of the a.c field, Eq (6) is expressed as

$$\Gamma = \Gamma_0 \sum_{k=-\infty}^{\infty} \frac{J_k^2(z)}{1+(k\omega\tau+z_c)^2}\left[1-\Delta^2(1-b^2)\sin(qd)\sum_{k=-\infty}^{\infty}1+(k\omega\tau+2z_c)^2\right.$$

$$\left.\times(2T\omega_q\sum_{k=-\infty}^{\infty}\frac{J_k^2(z)}{1+(k\omega\tau+z_0)^2})^{-1}\right] \quad (7)$$

**Discussion and Conclusion**

The absorption coefficient of the acoustic wave in the presence of an external field is calculated as in Eq (7) and the dependence of $\Gamma/\Gamma_0$ on $z_c$ is presented graphically at different values of $z$ and $\omega\tau$ (see(fig 1 - 8)).

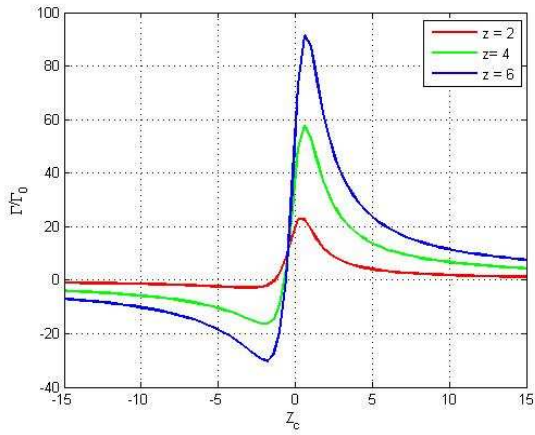

Fig1: Dependence of $\Gamma/\Gamma_0$ on $z_c$ for $\omega\tau \ll 1$ for varying $z$

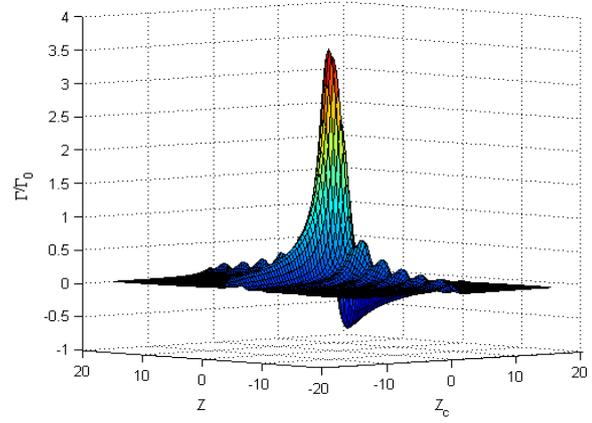

Fig 2: Dependence of $\Gamma/\Gamma_0$ on $z_c$ and $z$ for $\omega\tau \ll 1$

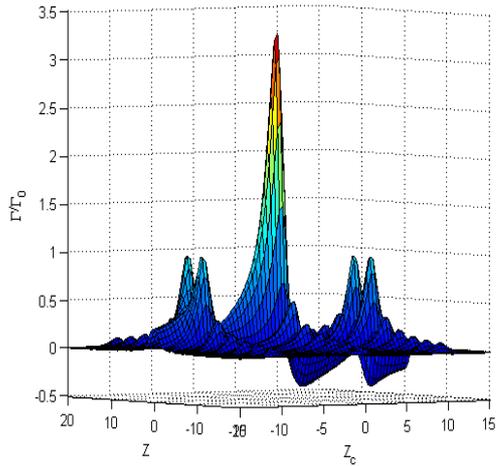

Fig3: Dependence of $\Gamma/\Gamma_0$ on $z_c$ and $z$ for $\omega\tau = 10$

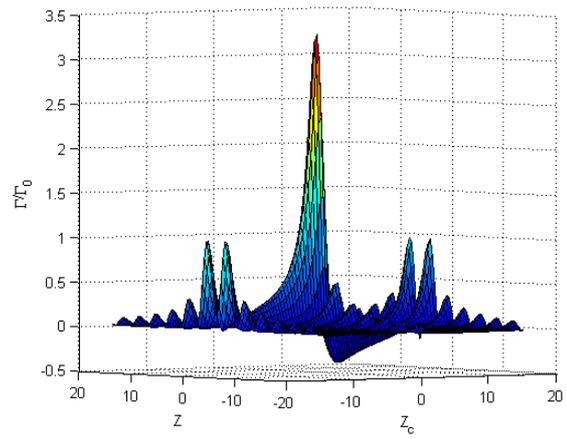

Fig. 4: Dependence of $\Gamma/\Gamma_0$ on $z_c$ and $z$ for $\omega\tau = 15$

The peak values of the graph increase by increasing $z$ and $\omega\tau$. The a.c field therefore is modulating the d.c field in a manner similar to the negative differential conductivity in *SL* where the a.c. field induces a synchronous modulation of the velocity with which the electrons traverse the Brillouin zone. In the above graphs (Fig 1), as $z_c$ and $\omega\tau$ increases, $\Gamma/\Gamma_0$ becomes more positive absorption until it reaches a critical value ($\omega \Rightarrow \omega_\beta$) where the graph switches over to the negative absorption side. The 3D graphs of $\Gamma/\Gamma_0$ on $z_c$ and $z$ is shown in Fig. 2, when $\omega\tau \ll 1$. With $\omega\tau \gg 1$, Photostimulated absorption is observed as in fig. 3 and 4 at $\omega\tau = 10 \text{ and } 15$ above. This observed shift in the peaks for different values of z and $\omega\tau$ causes an asymmetric dispersive curve which favors more absorption than amplification. Studying the behavior of the graph in the range

$\left[\dfrac{-\pi}{d}, \dfrac{2\pi}{d}\right]$ reveals an interesting phenomenon see (fig. 5 ).

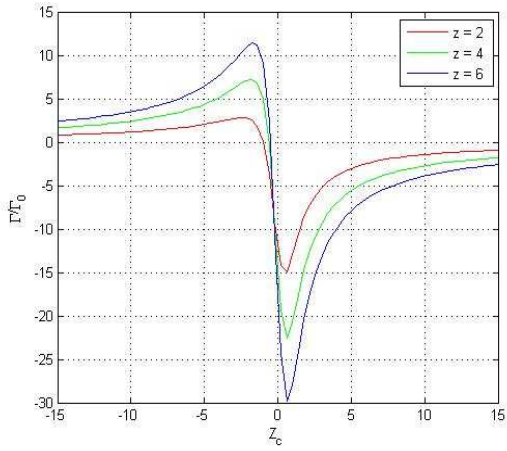

Fig 5: Dependence of $\Gamma/\Gamma_0$ on $z_c$ for $\omega\tau \ll 1$

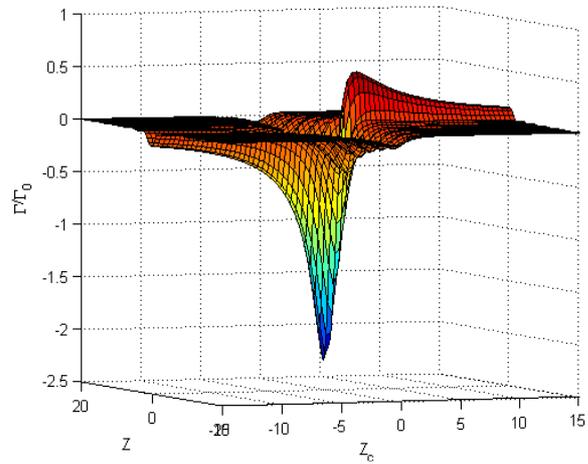

Fig 6: Dependence of $\Gamma/\Gamma_0$ on $z_c$ and $z$ for $\omega\tau \ll 1$.

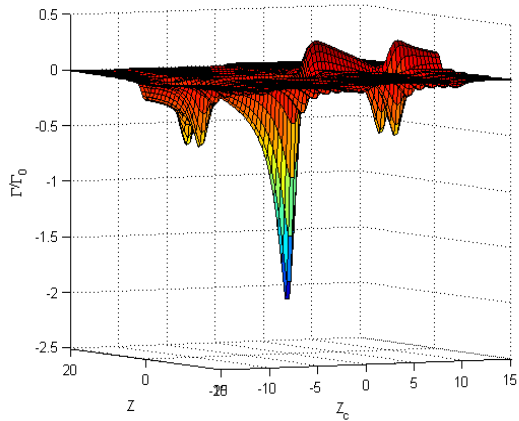

Fig 7: Dependence of $\Gamma/\Gamma_0$ on $z_c$ and $z$ for $\omega\tau = 10$

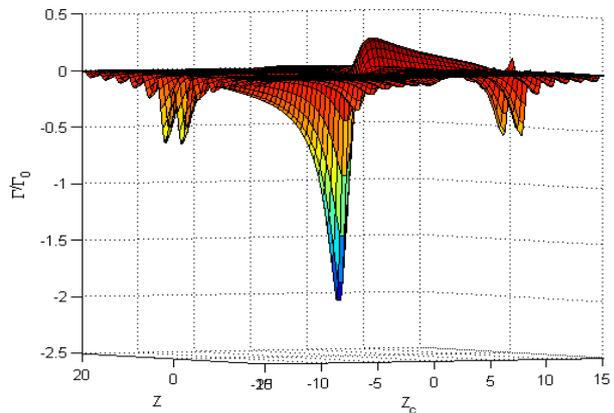

Fig 8: Dependence of $\Gamma/\Gamma_0$ on $z_c$ and $z$ for $\omega\tau = 15$

The following were assumed for numerical estimates of $E_0^{SL}$, $T = 300K$, $\Delta = 0.1 eV$, $d = 10^{-8} cm$, $s = 5*10^5 cms^{-1}$, $\tau = 10^{-12}$, and $\omega_q = 10^{12} s^{-1}$. It is observed from the graph that there is an inversion and amplification far exceeds absorption and the ratio $\frac{|\Gamma/\Gamma_0|_{min}}{|\Gamma/\Gamma_0|_{max}} \approx 3$. This indicates an amplification of the phonons generated in the terahertz range leading to the possibility of obtaining a hypersound MASER. Fig. 6 shows a 3D graph of Eq (7) by considering $\Gamma/\Gamma_0$ against $z$ and $z_c$. Photostimulated amplification is reported in the same range (see (fig. 7 and 8)).

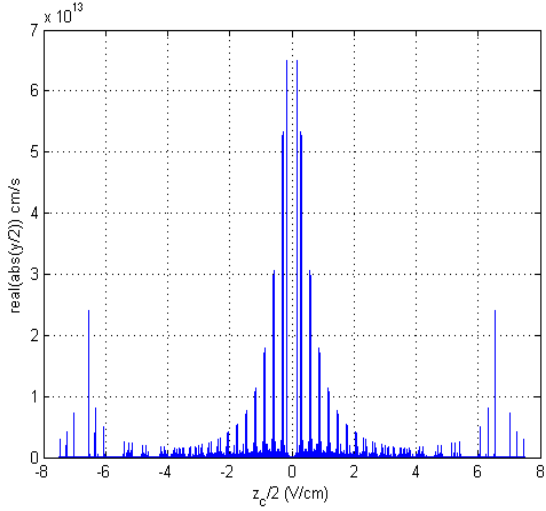 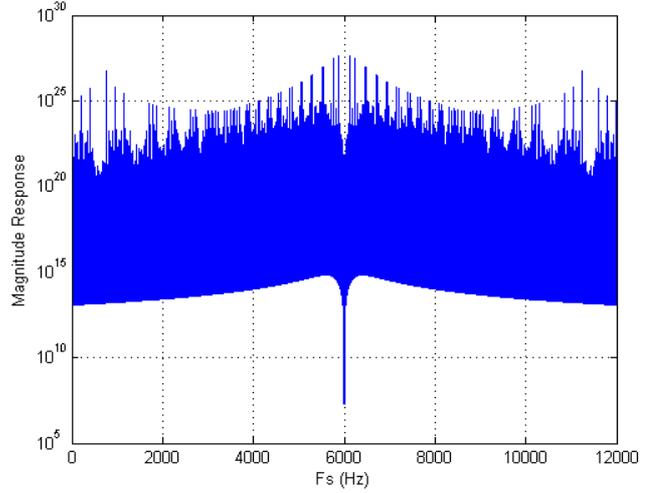

Fig 9: Buildup of acoustic energy from noise (Phonon spectrum) at Fs =12e3

Fig 10: The power spectrum

Fig (11) is the spectrum of acoustic noise (phonon spectrum) obtained by analyzing the useful part of the spectrum as

$$y = real(abs(y/2)) \qquad (8)$$

The use of the superlattice as a phonon filter for the generation hypersound amplification can be achieved by varying the sampling frequency to attain different phonon spectrum and their corresponding power spectrum as in fig. 10. The snapshot of the animated form of the various graphs, (see fig. 11)) reveals an interesting phenomena that is the growing of the hypersound MASER which is represented by the inverted twin peaks.

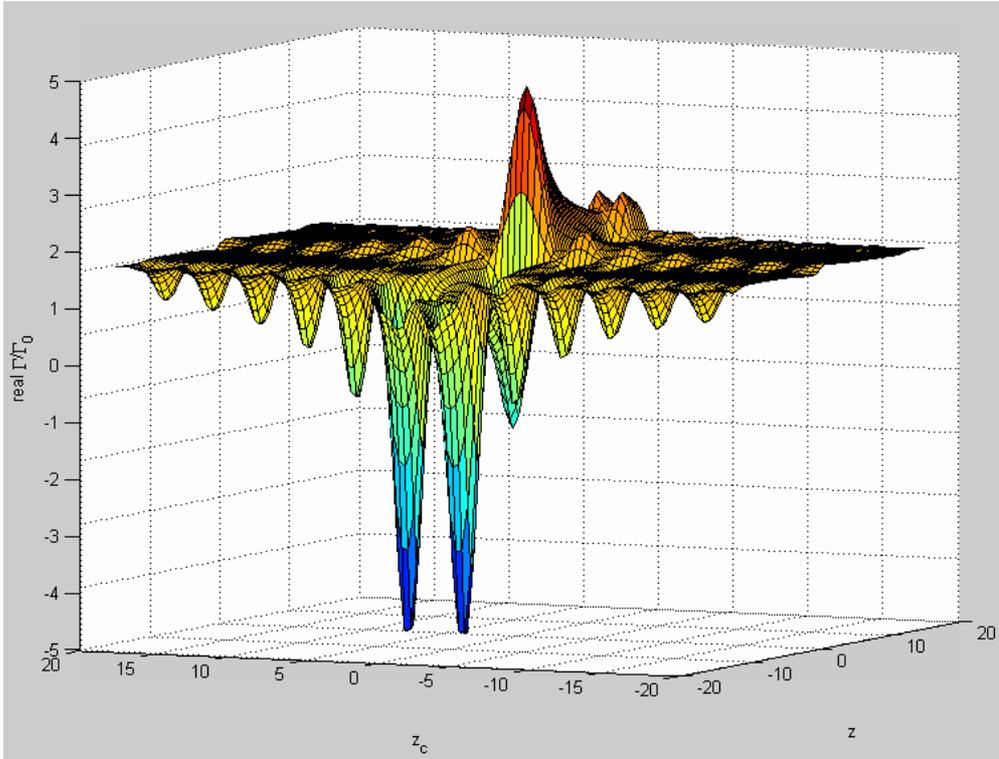
fig 11: A snap shot of an animated graph showing a twin peak of acoustic MASER